\documentclass[floatfix, preprintnumbers, amsmath, amssymb, 10pt]{revtex4}
\usepackage{graphicx}
\usepackage{dcolumn}
\usepackage{bm}

\def\Dsl{\hbox{/\kern-.6700em\it D}} % D slash
\def\dsl{\hbox{/\kern-.5300em$\partial$}}

\def\eqa{\begin{eqnarray}}
\def\eeqa{\end{eqnarray}}
\def\eq{\begin{equation}}
\def\eeq{\end{equation}}
\def\be{\begin{equation}}
\def\ee{\end{equation}}
\def\bea{\begin{eqnarray}}
\def\eea{\end{eqnarray}}
\def\nn{\nonumber}

\def\eq{Eq.\,}

\begin{document}

\date{\today}

\title{The Effects of Gravitational Back-Reaction on Cosmological Perturbations}

\author{Patrick Martineau $^{1)}$} \email[email: ]{martineau@hep.physics.mcgill.ca}
\author{Robert H. Brandenberger$^{1,2)}$} \email[email: ]{rhb@het.brown.edu}

\affiliation{1) Dept. of Physics, McGill University, 3600 University Street,
Montr\'eal QC, H3A 2T8, Canada}
\affiliation{2) Dept. of Physics, Brown University, 
Providence R.I. 02912, U.S.A.} 

\begin{abstract}

Because of the nonlinearity of the Einstein equations, the cosmological
fluctuations which are generated during inflation on a wide range of wavelengths
do not evolve independently. In particular, to second order in perturbation
theory, the first order fluctuations back-react both on the background geometry
and on the perturbations themselves. In this paper, the gravitational 
back-reaction of long wavelength (super-Hubble) scalar metric fluctuations on the  
perturbations themselves is investigated for a large class of inflationary models. 
Specifically, the equations describing the evolution of long wavelength 
cosmological metric and matter perturbations 
in an inflationary universe are solved to second order in both the amplitude of the 
perturbations and in the slow roll expansion parameter. Assuming that the linear
fluctuations have random phases, we show that the fractional correction to
the power spectrum due to the leading infrared back-reaction terms does not change
the shape of the spectrum. The amplitude of the effect is suppressed by the
product of the inflationary slow-roll parameter and the amplitude of the linear
power spectrum. 
%The effect of back-reaction
%is found to increase as the spectral index of the fluctuations approaches
%that of scale-invariance and can lead to observable effects for a spectrum
%which is sufficiently close to scale-invariant. Secondly, given an initial spectral 
%tilt at first order, second order effects generically modify the spectrum of 
%perturbations, leading to a greater tilt. 
The non-gaussianity of the spectrum 
induced by back-reaction is commented upon. 

\end{abstract}

\maketitle
       
%
%\leftline{hep-th/} } 

\section{Introduction}
The study of cosmological fluctuations is one of the cornerstones of modern cosmology. 
In order for a cosmological model to be considered successful, it must be able to 
reproduce, among other things, the power spectrum of the perturbations. These perturbations 
leave their mark as anisotropies in the cosmic microwave background (CMB) and go on to act 
as seeds for structure formation. The theory of cosmological perturbations
establishes the bridge between observations (namely observations of fluctuations in the 
CMB and in the distribution of structure in the universe) and the
physics of the very early universe which is responsible for providing the generation
mechanism for the fluctuations.

Arguably, the most successful cosmological paradigm for the early universe is the 
inflationary scenario \cite{Guth}. Inflationary cosmology provides a
mechanism for generating the primordial fluctuations and predicts a nearly 
scale-invariant spectrum of Gaussian, adiabatic, scalar fluctuations 
\cite{Mukhanov} (see also \cite{Press,Lukash,Sato}). These predictions 
have been verified experimentally to high accuracy in the last decade (see, e.g., the
most recent WMAP CMB anisotropy maps \cite{WMAP}). Since, in addition, inflationary
cosmology solves numerous problems of Standard Big Bang cosmology \cite{Guth} (for a 
review of these problems along with a discussion of some conceptual problems  
inherent to inflation see \cite{RHBKish}), inflation is 
widely considered as a key ingredient of the theory of the early universe.

At the present time, however, inflationary cosmology does not quite have the status of a 
{\it theory}. It is best thought of as a 
successful scenario that resolves many of the problems that plague Big Bang cosmology. There 
are a large number of different {\it models} that result in accelerated (i.e. inflationary) 
expansion. However, many models share a common feature in 
that they involve a scalar field (the ``inflaton") which undergoes a period in which it 
rolls slowly down it's potential - leading to what is known as ``slow-roll inflation".
Our analysis will be in the context of a general slow roll inflation model.

The theory of cosmological perturbations (see e.g. \cite{MFB} for a 
comprehensive review, and \cite{RHBrev} for a recent abbreviated overview), a formalism 
crucial to the understanding of CMB anisotropies, is usually studied within the framework 
of linearized gravity. One writes down an ansatz for the form of the perturbed metric about 
a homogeneous and isotropic background space-time, 
linearizes the Einstein equations in the amplitude of the perturbations, and 
solves the resulting equations. In this scheme, all Fourier modes of the fluctuations
evolve independently, as must be the case in a linear approximation. The Einstein equations
which govern the evolution of space-time and matter are, however, nonlinear.
Thus, retaining terms quadratic and higher in the perturbation amplitude leads to 
interactions between different perturbation modes. These interactions determine the 
``gravitational back-reaction", the difference between the full evolution of the space-time
and what would be obtained in linear theory, and will lead to 
potentially important modifications of the results obtained at linear order. In
particular, they may effect the key qualitative predictions of inflation, namely the scale 
invariance of the spectrum and its Gaussianity.

The period of inflation in most scalar field-driven inflationary models is very long
(measured in units of the Hubble time during inflation). Thus, the red-shifting of scales
leads to the population of a large phase space of long wavelength modes (modes with a
wavelength larger than the Hubble radius). The back-reaction of such long wavelength
modes on the background space-time was first studied in \cite{MAB,ABM} (making use
of the concept of an effective energy-momentum tensor of fluctuations first used in studies
of short wavelength gravitational waves \cite{Hartle}). It was found that this
effective energy-momentum tensor acts like a negative cosmological constant with a
magnitude which increases in time as the phase space of long wavelength modes grows.
A physical explanation of this effect in the quasi-homogeneous approximation to the
evolution equations was recently provided in \cite{Harry}. The effect can become
non-perturbatively large \cite{ABM} if the period of inflation is sufficiently long
and leads to a change in the Hubble expansion rate. This change is physically measurable
in models with at least two matter fields \cite{Ghazal2}. In models with only one
matter field, however, the leading infrared back-reaction terms are not physically
measurable by a local observer \cite{Unruh,Ghazal1,AW} (see also \cite{Afshordi}, and
see \cite{RHBrev2} for a review of previous work on gravitational back-reaction in
inflationary cosmology \footnote{There has been a lot of recent interest in the
possibility that the leading gradient terms of long-wavelength modes might back-react
on local observables in a way that mimics dark energy \cite{Kolb} (see also
\cite{GhazalNiayesh}). However, objections to this possibility have been
raised \cite{ACG,Flanagan,Seljak,Rasanen,Wiltshire}.}).

Given that the back-reaction of first order cosmological perturbations on the background
cosmology can become large, it is important to determine whether this gravitational 
back-reaction can also lead to large effects on the fluctuations themselves, potentially
also changing their key characteristics like almost scale-invariance and Gaussianity.

It is the intent of this article to solve the perturbed Einstein equations to quadratic 
order, and determine the modifications to the results of the linear analysis, focusing
on the effects on long-wavelength fluctuations. There
has been a significant body of previous work devoted to second order cosmological
perturbations, see e.g. \cite{Bruni,Bartolo,LythWands,WandsMalik,Hwang,Maldacena} and
papers quoted therein. The effect of long-wavelength modes has also been studied in
the ``separate universe'' approach \cite{Salopek,Rigopoulos,Afshordi}, in which the
effect of long-wavelength perturbations in encoded as a change in the background
geometry. Our approach is similar. In particular, we will neglect spatial
gradients in the equations of motion, and thus are focusing on the leading
infrared contributions to back-reaction. What differentiates our 
analysis from previous work is the emphasis on
the fact that, in an inflationary universe, modes continuously move into the
infrared sector (wavelengths greater than the Hubble radius), and that thus the 
infrared phase space grows. This leads to the concern that back-reaction effects
grow without limits, a concern which is the main motivation for our work.  

The main result of our study, however, is that the leading infrared contributions
of back-reaction to the power spectrum of cosmological fluctuations is very small.
Assuming that the linear fluctuations have random phases, as they do in the
simplest inflationary models, the relative contribution of our back-reaction terms
to the power spectrum is suppressed by the product of a dimensionless inflationary
slow-roll parameter and the amplitude of the linear density fluctuations.

We free 
ourselves of constraints imposed by any specific inflationary model by abstaining from 
picking a specific form of the inflaton potential. Rather, we assume that the inflationary 
slow-roll conditions are satisfied and we employ a very general form of the potential, one 
that allows us to easily interpret our results within the context of an explicit slow-roll 
realization. Thus, with regards to cosmological models, our approach is extremely general, 
while still retaining its ability to be specific.

Section 2 of this paper is concerned with the general setup of the calculation. Here, the 
background equations are solved to second order in the slow-roll parameter ($\epsilon$),
neglecting spatial gradients, and 
the corrections to both the background scalar and the Hubble constant are determined, again, 
to second order in $\epsilon$.
Section 3 begins with a brief review of classical, relativistic, cosmological perturbation 
theory. The linear perturbation equations are solved using the background obtained in the 
previous section.
The non-linear equations make their appearance in Section 4, where the next-to-leading 
order terms (those due to back-reaction) are solved for.
Section 5 interprets these results as modifications to the linear terms, and their effect 
on the spectrum of perturbations is determined.
Finally, the effects of back-reaction on the Gaussianity of the perturbations is 
investigated in Section 6.

\section{Background}

Before solving for the perturbations, we must determine the appropriate background for our 
model. We confine our attention to a model consisting of pure gravity (with vanishing 
cosmological constant)
and a single, homogeneous, scalar field $\phi$, which we presumes satisfies the 
slow-roll conditions. We write
\be \label{fieldansatz}
\phi(x,t) \, = \, \phi_{0}+\epsilon f_{1}(t)+\epsilon^{2} f_{2}(t) \, ,
\ee
where $\phi_{0}$ is a constant and $\epsilon$ is a dimensionless slow-roll parameter 
whose value depends on the specific model in question.
In consequence of (\ref{fieldansatz}), our background space-time will be approximately 
de Sitter space,
\be
a(t)=e^{H(t)t} \, .
\ee
Here, the Hubble rate $H(t)$ is slowly time-dependent and can be expanded as
\be
H(t)=H_{0}+\epsilon h_{1}(t)+ \epsilon^{2} h_{2}(t) \, ,
\ee
where $H_{0}$ a constant.

The dynamics of the scalar field is determined by its potential $V(\phi)$. 
In order not to limit our results to any specific model, but, rather, to show them to be 
a generic feature of slow-roll inflation, we use the following expansion (related to
a power series expansion of the potential about the field value $\phi_0$):
\be \label{potexp}
V(\phi) \, = \, \mu\phi_{0}+\epsilon\lambda\phi(t)+\epsilon^{2}\Lambda\phi^{2}(t) \, 
\equiv \, V^{(0)} + V^{(1)} + V^{(2)} \, ,
\ee
where superscripts indicate the order in $\epsilon$ and 
\bea
\mu \phi_0 &=& V(\phi_0) - \phi_0 V^{'}(\phi_0) + {1 \over 2} \phi_0^2 V^{''}(\phi_0) \, , \\
\epsilon \lambda &=& V^{'}(\phi_0) - \phi_0 V^{''}(\phi_0) \, \\
\epsilon^2 \Lambda &=& {1 \over 2} V^{''}(\phi_0) \, ,
\eea
the primes denoting a derivative with respect to the field. Thus, we see that
$\mu$,$\lambda$, and $\Lambda$ are dimensionful parameters depending entirely on the 
form of the potential in the neighborhood of $\phi_0$. In essence, this represents the 
series expansion (in ${\epsilon}$) of any potential that can be used to generate slow-roll 
inflation.

In light of this, we can use the Klein-Gordon 
\be \label{KGeq}
\ddot\phi+3\frac{\dot a}{a}\dot\phi+\frac{\partial V}{\partial \phi}=0 \, ,
\ee
and the Friedmann equations   
\bea
\-3 \dot a^{2} + 4\pi a^{2} \dot{\phi}^{2} + 8\pi a^2V \, &=& \, 0 \, ,\\
\frac{\dot a^{2}}{a^{2}} + 2\frac{\ddot a}{a} + 4 \pi \dot \phi^{2}-8\pi V &=& 0
\eea
(solved order by order in $\epsilon$)
to solve for the background, accurate to second order in the slow roll parameter.

We find that
\be
H^{2}_{0} \, = \, 8\pi\mu\phi_{0} \, ,
\ee
\be
f_{1}(t) \, = \, \frac{\lambda}{3H_{0}}(\frac{1}{3H_{0}} - t - 
\frac{e^{-3H_{0}t}}{3H_{0}}) \, , 
\ee
\be
f_{2}(t) \, = \, \frac{\lambda^{2}}{18 H^{2}_{0}\mu}(-2+3H_{0}t + 
e^{-3H_{0}t}(3H_{0}t+2)) \, ,
\ee
\be
h_{1}(t) \, = \, \lambda\sqrt{\frac{2\phi_{0}}{3\mu\pi}} \, ,
\ee
and
\bea
h_{2}(t) = &&-\frac{1}{1296\pi\mu^{2}H_{0}^{4}t}(288H^{2}_{0}\lambda^{2}t^{2}\pi^{2}\mu^{2}
+ 162H^{5}_{0}\lambda^{2}t\pi-243\Lambda H^{7}_{0}t\\ \nn
&&+16\lambda^{2}\pi^{2}\mu^{2}e^{-6H_{0}t}-128\lambda^{2}\pi^{2}\mu^{2}e^{-3H_{0}t}-
288\lambda^{2}\pi^{2}\mu^{2}H_{0}t+112\lambda^{2}\pi^{2}\mu^{2}) \, .
\eea

Note that the corrections to the scalar field have the property that they and their 
first temporal derivatives vanish at $t=0$, which coincides with the onset of slow-rolling. 
Despite the overall factor of $t^{-1}$ preceding $h_{2}(t)$, this term is not singular at 
the origin, as can be seen by expanding in a power series about $t=0$.

\section{Linearized Theory}

The theory of cosmological perturbations is usually restricted to an analysis of the 
linearized (in the amplitude of the perturbations) Einstein equations. In this section, 
we review the salient results (see e.g. \cite{MFB,RHBrev} for comprehensive reviews).

The most general perturbed line element can be written as
\be
ds^{2} \, = \, a^{2}(\eta)[(1+2\Phi)d\eta^{2} - 2(B_{;i}+S_{i})dx^{i}d\eta -
[(1-2\Psi)\gamma_{ij} + 2E_{;ij}+(F_{i;j}+F_{j;i})+h_{ij}]dx^{i}dx^{j}]
\ee
where $\Phi,\Psi,B,E$ represent scalar, $S_{i},F_{i}$ vector, and $h_{ij}$ tensor 
metric perturbations, respectively. These are distinguished by their transformation 
properties under three dimensional rotations. For inflationary cosmology
and in linear perturbation theory, we can
discard vector modes since they do not grow, and tensor modes because they grow
at a slower rate than scalar metric fluctuations. Hence, we focus on the scalars. It is
possible to choose a gauge in which $E = B = 0$, the so-called 
`longitudinal' or `conformal Newtonian' gauge. This
gauge is convenient for calculational purposes. For matter without anisotropic
stresses to linear order in the matter field perturbations, it follows from the
off-diagonal space-space Einstein equations that $\Phi = \Psi$. Hence, for such
matter the metric in longitudinal gauge takes the form
\be \label{longitud}
ds^{2} \, = \, (1+2\kappa\Phi(\vec{x},t))dt^{2} - 
a^{2}(t)(1-2\kappa\Phi(\vec{x},t))(dx^{2}+dy^{2}+dz^{2}) \, .
\ee
In the above, $\kappa$ is a dimensionless parameter which indicates the order
of the term in gravitational perturbation theory.

At second order, the perturbed metric is in general much more complicated. In
particular, there is mixing between scalar, vector and tensor modes, and an
anisotropic stress in generated, leading to $\Phi \neq \Psi$. However, it can
be seen explicitly that all of the complicating terms contain spatial gradients
and can hence be neglected in our study of the back-reaction effects of
long wavelength modes. Thus, if we work to leading order in the infrared
terms, we can neglect vector, tensor modes and anisotropic stress. Thus,
we can apply longitudinal gauge also at second order. We will use this
gauge in the following.

To incorporate effects due to slow-rolling, we 
expand $\Phi$ in powers of $\epsilon$ to obtain
\be
\Phi_1(\vec{x},t) \, = \, \Psi_{1}(\vec{x},t) +\epsilon\alpha_{1}(\vec{x},t) +
\epsilon^{2}\beta_{1}(\vec{x},t) \, .
\ee
The subscript $_{1}$ denotes the fact that the effects are linear in $\kappa$.

In order to source our first order metric perturbations, matter perturbations 
$\delta_{1}(\vec{x},t)$ must be present, i.e.
\be
\phi(\vec{x},t)  \longrightarrow \phi(t) + \kappa\delta_{1}(\vec{x},t) \, ,
\ee
where $\phi(t)$ is the background matter solution given by (\ref{fieldansatz})

At the linear level, perturbations decouple in Fourier space. It is thus
convenient to track the evolution of each Fourier mode individually. At higher
orders in perturbation theory, there will be mixing between the modes.
Typically, cosmological perturbations are classified into two distinct sectors: 
sub-Hubble (UV) and super-Hubble (IR) modes. During inflation, the phase space 
associated with super-Hubble modes grows exponentially, while that of the 
sub-Hubble modes remains constant. In addition, the modes which are 
important from the point of view of structure formation and CMB anisotropies
are super-Hubble during the last 50 e-foldings of inflation. 
It is for these reasons that we focus our attention on these modes and ignore their 
UV counterparts. This choice justifies our dropping of spatial derivatives in the 
equations of motion. 

Using the standard result for the energy-momentum tensor of a scalar field 
\be
T_{\mu\nu} \, = \, \phi_{;\mu}\phi_{;\nu} - 
g_{\mu\nu}(\frac{\phi^{;\alpha}\phi_{;\alpha}}{2}-V(\phi)) \, ,
\ee
expanding the Einstein equations in a power series in $\kappa$, and truncating after 
first order leads to the equations of motion for scalar cosmological perturbations, 
which read (see e.g. \cite{MFB})
\bea
(ii):\,\,\,\,\, && -\frac{\dot{a}}{a}\dot{\Phi_{1}} + 8\pi \dot{\phi} \dot{\delta_{1}}
- 2\ddot{\Phi_{1}} - 8\frac{\ddot{a}}{a}\Phi_{1} - 4\Phi_{1}\frac{\dot{a}^{2}}{a^{2}}
- 16\pi\Phi_{1}(\dot{\phi})^{2}\\ \nn
&&- 8\pi V^{(1)}(\phi) + 16\pi V^{(0)}(\phi) \Phi_{1} \, = \, 0,\\ 
(00):\,\,\,\,\, && 6\dot{\Phi_{1}}\frac{\dot{a}}{a} + 8\pi \dot{\phi} \dot{\delta_{1}}
+ 8\pi V^{(1)}(\phi) + 16\pi V^{(0)}(\phi) \Phi_{1} \, = \, 0 \, ,
\eea
where the overdot denotes a derivative with respect to cosmic time, and the 
$(ii)$ and $(00)$ indicate the tensor indices of the Einstein equations in question.

These equations can be solved to yield
\bea 
\Psi_{1}(\vec{x},t) \, &=& \, 0 \, , \\
\alpha_{1}(\vec{x},t) \, &=& \, \psi_{1}\int{d^{3}\vec{k} 
f(\vec{k}) e^{i\vec{k}\cdot\vec{x}} e^{i\alpha_{\vec{k}}}} V^{1/2} \, \label{alpha1} \\
\beta_{1}(\vec{x},t) \, &=& \, -\psi_{1}\frac{2\Lambda\mu\phi_{0} - \lambda^{2}}{\lambda\mu}
(-1+e^{-H_{0}t})\int{d^{3}\vec{k}f(\vec{k}) e^{i\vec{k}\cdot\vec{x}}
e^{i\alpha_{\vec{k}}}} V^{1/2} \, \\
\delta_{1}(\vec{x},t) \, &=& \, -2\frac{\mu\phi_{0}\psi_{1}}{\lambda}
\int{d^{3}\vec{k}f(\vec{k}) e^{i\vec{k}\cdot\vec{x}} e^{i\alpha_{\vec{k}}}} V^{1/2} \, ,
\eea
where $\psi_{1}$ a constant representing the amplitude of the spectrum
of linear fluctuations, $f(\vec{k})$ is the function describing the shape of
the spectrum, $\alpha_{\vec{k}}$ are the phases (assumed later on to be
random), and $V$ is the cutoff volume used in the definition of the
Fourier transform. Note that $\vec{k}$ denotes the co-moving momentum. We have
introduced the cutoff volume in the definition of the Fourier transform in order
that the dimension of the Fourier transform $\tilde \Phi(\vec{k})$ of $\Phi(\vec{x})$
is $k^{3/2}$, which in turn ensures that there is no volume arising in the
relation between the power spectrum $P_{\Phi}(k)$ and $\tilde \Phi(\vec{k})$
(see Section 5).

Let us comment briefly on the interpretation of this solution. Note first that
the fact that $\Psi_{1}$ vanishes is a consistency check on our analysis. In the pure
de Sitter limit (no rolling of the scalar field), there are no scalar metric fluctuations
to first order in perturbation theory (in $\kappa$). Since the equation of
motion is second order, there are two fundamental solutions for each mode. In the 
long wavelength (super-Hubble) limit, the dominant solution is constant in time, and
the sub-dominant solution is a decaying mode. We see that $\alpha_{1}$ is a
the constant perturbation sourced by $\delta_{1}$, while $\beta_{1}$
is generated by the rolling of the scalar field and is a combination of the constant
and decaying mode.

\section{Back-Reaction}

To second order in gravitational perturbation theory (expansion in $\kappa$), there
are interactions between the Fourier modes of the fluctuation variables caused by
the non-linearities of the Einstein equations. 
In order to determine the effects of non-linearities, we make the substitutions for
the metric perturbation variable
\be
\Phi(\vec{x},t) \longrightarrow \Phi_1(\vec{x},t) + \kappa \Phi_{2}(\vec{x},t) \, , 
\ee
where $\Phi_2$ can be in turn expanded in terms of the slow-roll parameter $\epsilon$ as 
\be
\Phi_{2}(\vec{x},t) \, = \, \Psi_{2}(\vec{x},t) + \epsilon\alpha_{2}(\vec{x},t)
+ \epsilon^{2}\beta_{2}(\vec{x},t) \, ,
\ee
and for the matter field
\be
\phi(\vec{x},t) \longrightarrow \phi(\vec{x},t) + \kappa^{2}\delta_{2}(\vec{x},t) \, .
\ee
Note that $\Phi_2$ and $\delta_{2}$ represent the effects of gravitational back-reaction.

In the following we will neglect spatial gradient terms in the equations of motion,
since we are interested in the infrared modes and in the coupling between different
infrared modes \footnote{It is also in this approximation that we can justify
writing the perturbed metric in the form (\ref{longitud}) - see e.g. \cite{Afshordi}
for a discussion of this point.}. To order $\kappa^{2}$ in the perturbative expansion, 
the Einstein equations then become
\bea 
(ii): \,\,\,\,\,&& 2 \ddot \Phi_{2} + 8\dot \Phi_{2} \frac{\dot a}{a} +
8\Phi_{2} (\frac{\ddot{a}}{a} + 2\pi \dot{\phi}^{2} + \frac{\dot{a}^{2}}{2a^{2}} -
2\pi V^{(2)}(\phi)) - 8\pi\dot {\phi}\dot{\delta_{2}}\,\, \\ 
\nn = &&\,\,4 \Phi_{1}(\ddot{\Phi_{1}} + 2\Phi_{1}\frac{\dot{a}^{2}}{a^{2}} +
6\dot{\Phi_{1}}\frac{\dot{a}}{a} + 4\Phi_{1}\frac{\ddot{a}}{a} +
8\pi\Phi_{1}\dot{\phi}\dot{\delta_{1}} + 4\pi V^{(1)}(\phi) \Phi_{1}, \\ 
(00): \,\,\,\,\,&& 6 \dot{\Phi_{2}}\frac{\dot{a}}{a} + 8\pi \dot{\phi} \dot{\delta_{2}} -
16\pi V^{(0)}(\phi) \Phi_{2}\,\, \\ 
\nn=&&\,\, 3 \dot{\Phi_{1}}^{2} - 4\pi\dot{\delta_{1}}^{2} - 
12\Phi_{1}\dot{\Phi_{1}}\frac{\dot{a}}{a} - 16\pi V^{(1)}(\phi) \Phi_{1} - 8\pi V^{(2)}(\phi).
\eea

These equations can be solved to yield
\bea
\Psi_{2}(\vec{x},t) \, &=& \, 0 \, , \\
\alpha_{2}(\vec{x},t) \, &=& \, 0 \, , \\
\beta_{2}(\vec{x},t) \, &=& \, \label{eq3} 
\psi^{2}_{1}\frac{(\Lambda\mu\phi_{0}-\lambda^{2})}{\lambda^{2}}
(-1+e^{-H_{0}t})g^{2}(\vec{x}) \, , \\
\delta_{2}(\vec{x},t) \, &=& \, 0 \, ,
\eea
with
\be
g(\vec{x}) \, = \, \int{d^{3}\vec{k} f(\vec{k}) e^{i\vec{k}\cdot\vec{x}}
e^{i\alpha_{\vec{k}}}} V^{1/2} \, .
\ee

Let us briefly comment on the physical interpretation of these results. First, the
vanishing of $\Psi_2$ is a consistency check since there are no scalar metric
fluctuations in pure de Sitter space. Since the linear fluctuations are first order
in $\epsilon$, they will only contribute to the second order perturbations to quadratic
order, and hence the vanishing of $\alpha_2$ is another consistency check on the
algebra. From the expression for $\Phi_{2}$ it is manifest that the second order 
perturbations are generated by the linear inhomogeneities 
$\Phi_{1}$ and $\delta_{1}$ at quadratic order. The vanishing of $\delta_2$ is
an interesting and unexpected result. It says that, to this order, there are no
back-reaction effects on the evolution of the background scalar field.

\section{Effects of Back-Reaction on the Power Spectrum}

Having now determined the form of the back-reaction terms, it is important to
estimate their amplitude. From observations of CMB anisotropies, we know that the
linear perturbations are of order $\kappa \psi_1 \epsilon \sim 10^{-5}$. The back-reaction
terms should be expected to be of order $(\kappa \psi_1 \epsilon)^{2}$. However, the
second order correction to a fixed Fourier mode receives contributions from all
linear Fourier modes to this order. Hence, one could expect the back-reaction effect
to be amplified by a phase space factor which measures the phase space of
Fourier modes which contribute. Since in inflationary cosmology, the phase space
of infrared modes is growing, and the linear fluctuations do not decrease in amplitude
on scales larger than the Hubble radius, the effects of back-reaction could be expected 
to grow in time and become non-perturbatively large. In this section we
show that, provided that the linear fluctuations have random phases, the
leading infrared quadratic back-reaction effects of linear fluctuations on the
power spectrum of $\Phi$ do not show any large phase-space enhancement. 
  
The total power spectrum $\mathcal{P}_{total}(\vec{k})$ of $\Phi$, 
including the leading infrared terms of second order in $\kappa$, 
can be written as
\bea 
\mathcal{P}_{total}(\vec{k}) \, && = \, k^3 |\tilde{\Phi_{\vec{k}}}|^2 \, = \,
\mathcal{P}_{1}(\vec{k})+\mathcal{P}_{2}(\vec{k}),\\ \nn
&&=\, |\tilde{\Phi}_{1}(\vec{k}) + \tilde{\Phi}_{2}(\vec{k})|^{2}k^{3},
\label{power1}
\eea
where $\tilde{\Phi_{i}}(\vec{k})$ are the Fourier transforms (using the definition of
Fourier transform including the cutoff volume as in Equation (\ref{alpha1})) 
of $\Phi_i(\vec{x})$. Making use of the results of Sections 3 and 4 we have
\bea
\tilde{\Phi}_{1}(\vec{k}) \, &=& \, \epsilon \psi_1 f(k) e^{i\alpha_{\vec{k}}}
\bigl[ 1 - \epsilon E(t) \bigr] \nn \\
\tilde{\Phi}_{2}(\vec{k}) \, &=& \, \epsilon^2 \psi_1^2 E(t) h(k) \, ,
\eea
where $h(k)$ is the Fourier transform of $g^2(x)$ and where we have introduced
the symbol $E(t)$ for the function
\be
E(t) \, \equiv \, 
{{2 \Lambda \mu \phi_0 - \lambda^2} \over {\lambda^2}} \bigl( -1 + e^{-H_0 t} \bigr)
\, .
\ee

The leading back-reaction correction to the power spectrum, denoted by
the function $\mathcal{P}_{BR}(\vec{k})$, comes from the cross
term in (\ref{power1}) and is thus linear in $\tilde{\Phi}_{2}(\vec{k})$:
\be
\mathcal{P}_{BR}(\vec{k}) \, = \, 2 \kappa 
|\tilde{\Phi}_{1}(\vec{k})\tilde{\Phi}_{2}(\vec{k})| k^3 \, .
\ee
Thus, the fractional correction to the power spectrum due to the leading
back-reaction contributions is
\bea
{{\mathcal{P}_{BR}(k)} \over {\mathcal{P}_{1}(k)}} \, &=& \,
2 \kappa {{\epsilon^2 \psi_1^2 E(t) h(k)} \over {\epsilon \psi_1 f(k) + 
\cal{O}(\epsilon^2)}} \nn \\
&\simeq& \, 2 \kappa \epsilon \psi_1 {{h(k)} \over {f(k)}} \, ,
\label{powerratio}
\eea
from which it follows that, modulo the ratio of $h(k)$ over $f(k)$, the 
back-reaction terms are suppressed, as expected, by $\kappa \epsilon \psi_1$.
The ratio of $h(k)$ over $f(k)$ is the possible large phase space
enhancement factor.

Before continuing, we specify the linear power spectrum. We choose a normalization
wavenumber $k_n$ and choose $\psi_1^2$ to be the amplitude of the power spectrum
at $k = k_n$. The function $f(k)$ describes the spectral shape. We choose a power
law with a tilt $\zeta$ away from scale-invariance, i.e. we write
\be \label{spectrum}
f(k) \, = \, \biggl({k \over {k_n}}\biggr)^{-3/2 - \zeta} k_n^{-3/2} \, .
\ee
It can easily be checked that $\mathcal{P}_{1}(k_n) = \psi_1^2$.

We now evaluate the magnitude of $h(k)$, assuming that the phases $\alpha_{\vec{k}}$
are random:
\bea
h(\vec{k}) \, &=& \, 
{1 \over {(2 \pi)^3}} V^{-1/2} \int{d^3x g^2(x) e^{-i \vec{k} \vec{x}}} \nn \\
&=& \, \int{d^3k_1 f(\vec{k}_1) f(\vec{k} - \vec{k}_1)
e^{i(\alpha_{\vec{k_1}} + \alpha_{\vec{k} - \vec{k}_1})}} V^{1/2} \, .
\label{hk}
\eea
Given that we are considering the effects of long-wavelength fluctuations,
we must restrict the above integral over $\vec{k}_1$ to run only over
super-Hubble modes, i.e.
\be
|\vec{k_1}| \, \leq \, H \, .
\ee
To estimate the magnitude of $h(\vec{k})$, we insert
the spectrum (\ref{spectrum}) into (\ref{hk}). If we consider the effects
of back-reaction on modes $\vec{k}$ which are sub-Hubble now, 
we can apply the approximation
\be \label{approx}
\biggl({{k - k_1} \over {k_n}}\biggr) \, \simeq \, {k \over {k_n}} \, ,
\ee
in which case the integral simplifies.

Assuming constant phases for the moment, the integral (\ref{hk}) can be
easily estimated
\be \label{result1}
h(k) \, \sim \, k^{-3/2 - \zeta} k_n^{2 \zeta} H^{3/2 - \zeta} V^{1/2} \, .
\ee
Note in particular from (\ref{result1}) that the k-dependence of 
$\mathcal{P}_{BR}(k)$ is the same as that of the linear power spectrum. The
leading effect of back-reaction thus does not change the power index of
the spectrum. However, for wavelengths close to the Hubble radius, the
approximation (\ref{approx}) is no longer good, and the correction terms
will yield changes to the index of the power spectrum. The second fact to
notice about the result (\ref{result1}) is the cutoff volume divergence.
This stems from the fact that as $V$ increases, more and more infrared modes
are contributing to the back-reaction. For constant phases, the effect
is additive. The volume divergence thus represents the phase space enhancement
which is the focus of this investigation.

Let us now consider the more realistic situation - realized in typical
inflationary models - in which the phases are random. A simple way
to estimate the effects of the random phases in (\ref{hk}) is to add up
the amplitudes of the back-reaction contributions of all infrared modes
$\vec{k}_1$ as a random walk. This means dividing the amplitude obtained
previously by $N(V)^{1/2}$, where $N(V)$ is the number of modes. Since
for a finite volume $V$ the wavenumbers are quantized in units of
$\Delta k \sim V^{-1/3}$, the number $N(V)$ becomes
\be
N(V) \, \sim \, \bigl({H \over {\Delta k}}\bigr)^3
\ee
in which case the result (\ref{hk}) becomes
\be \label{result2}
h(k) \, \sim \, k^{-3/2 - \zeta} k_n^{2 \zeta} H^{- \zeta} \, .
\ee
Inserting this into (\ref{powerratio}), we obtain our final result
\be \label{result3} 
{{\mathcal{P}_{BR}(k)} \over {\mathcal{P}_{1}(k)}} \, \sim \,
2 \kappa \epsilon \psi_1 \biggl({{k_n} \over H} \biggr)^{\zeta} \, .
\ee

The main conclusion we draw from (\ref{result3}) is that there is
no phase space enhancement of the back-reaction of long wavelength
modes on the spectrum of cosmological perturbations, in contrast to
the positive enhancement found for the back-reaction on the background
metric. Given the absence of such a phase space enhancement, we find - as
expected - that the back-reaction terms in the power spectrum are suppressed
compared to the terms coming from the linear perturbations by
$\kappa \epsilon \psi_1$. Thus, they are completely negligible  in the
case of a COBE-normalized spectrum of almost scale-invariant linear
fluctuations. In addition, we find that the leading back-reaction
terms do not change the spectral index.

\section {Non-Gaussianity of the Spectrum Due to Back-Reaction}

Having established that the back-reaction of infrared modes cannot
substantially modify the amplitude and spectral tilt of the power spectrum 
of perturbations, we make some comments regarding 
the effects of back-reaction on the Gaussianity of the spectrum.

The inclusion of higher-order terms implies correlations between different modes, 
thus breaking strict Gaussianity. However, the question remains: how badly broken is 
it? To estimate this, we turn our attention to the bispectrum (three-point function). 

In the case of purely Gaussian distribution, all odd moments are identically zero. 
Therefore, the non-vanishing of the bispectrum indicates that the distribution cannot 
be Gaussian.

We take it for granted that the bispectrum does not vanish (for examples of the 
three-point function see \cite{Bartolo2,Maldacena}. 
In the context of higher order perturbation theory (see e.g. \cite{Bartolo}), 
however, its amplitude is exceedingly small. We estimate it to be no larger than 
of order $\kappa^{4}\epsilon$, thus making it quite unlikely to be detected 
experimentally. Thus, we conclude that, although back-reaction modifies the 
distribution, Gaussianity remains an excellent approximation.

\section {Conclusions}

In this paper, we have studied the back-reaction of long wavelength 
linear fluctuations on the power spectrum, 
produced by the mode mixing which occurs as a 
consequence of the non-linearity of the Einstein equations. We find
that, assuming that the phases of the linear fluctuations are random,
there is no phase space enhancement of the back-reaction effect. The
leading infrared
back-reaction contributions are suppressed by $\kappa \epsilon \psi_1$
compared to the contribution of the linear fluctuations, where 
$\kappa \psi_1$ is a measure of the amplitude of the linearized metric
fluctuations, and $\epsilon$ is an inflationary slow-roll parameter.
These leading back-reaction terms do not modify the tilt of the power
spectrum on scales substantially smaller than the Hubble radius. 
Note that in the case of correlated phases of the linear
fluctuations, a much larger back-reaction effect is possible.

We have also seen that the modifications to the Gaussianity of the leading order 
perturbations are negligible. The small size of these modifications 
goes a long way towards justifying the linear approximation to cosmological 
perturbation theory. However, our results do not exclude the possibility that
large amplitude local fluctuations can effect the measured fluctuations, as
very recently suggested in \cite{Tomita2} (based on the second order formalism
developed in \cite{Tomita1}) \footnote{The possibility that local fluctuations
can have a measurable effect on background quantities such as the deceleration
parameter has recently been suggested in \cite{Rasanen2,Notari}.}.

Our work differs from previous work on second order fluctuations in that it
emphasizes the fact that, in an accelerating universe, the phase space of super-Hubble
modes is increasing in time. In contrast to what occurs in the case of the back-reaction
on the homogeneous mode, the back-reaction on the fluctuating modes themselves does
not increase without limits as a function of time. Compared to previous analyses, our
work also gives an easier way to derive the leading-order
effects of long-wavelength cosmological fluctuations.
Our results have been derived in the context of an arbitrary slow-roll inflationary 
model and are thus valid for a wide range of cosmological scenarios.
 
\acknowledgments{This work is supported by funds from McGill University,
by an NSERC Discovery Grant (at McGill) and
(at Brown) by the US Department of Energy under Contract DE-FG02-91ER40688, 
TASK A.}

\end{document}